# TOWARDS THE ONTOLOGY WEB SEARCH ENGINE

Olegs Verhodubs

oleg.verhodub@inbox.lv

**Abstract.** The project of the Ontology Web Search Engine is presented in this paper. The main purpose of this paper is to develop such a project that can be easily implemented. Ontology Web Search Engine is software to look for and index ontologies in the Web. OWL (Web Ontology Languages) ontologies are meant, and they are necessary for the functioning of the SWES (Semantic Web Expert System). SWES is an expert system that will use found ontologies from the Web, generating rules from them, and will supplement its knowledge base with these generated rules. It is expected that the SWES will serve as a universal expert system for the average user.

**Keywords:** Ontology Web Search Engine, Search Engine, Crawler, Indexer, Semantic Web

I. INTRODUCTION

The technological development of the Web during the last few decades has provided us with more information than we can comprehend or manage effectively [1]. Typical uses of the Web involve seeking and making use of information, searching for and getting in touch with other people, reviewing catalogs of online stores and ordering products by filling out forms, and viewing adult material. Keyword-based search engines such as YAHOO, GOOGLE and others are the main tools for using the Web, and they provide with links to relevant pages in the Web. Despite improvements in search engine technology, the difficulties remain essentially the same [2]. Firstly, relevant pages, retrieved by search engines, are useless, if they are distributed among a large number of mildly relevant or irrelevant pages. Secondly, relevant pages may not be retrieved by search engines at all. In truth, it is a rare phenomenon that happens with modern search engines. Thirdly, the pages, retrieved by search engines, are sensitive to vocabulary. Initial keywords do not provide with the results we want, because relevant pages use different terminology from the original query. Finally, retrieved results are single web pages and hence it is necessary manually extract the partial information and put it together [2]. This requires the person, who will browse retrieved pages and extract the information he is looking for. The need of a new approach to manage information is beyond doubt. Development of a SWES (Semantic Web Expert System) is an attempt to change this situation for the better, and this development is the main goal of the research. The SWES is a new expert system, which is based on the Semantic Web technologies [3]. It is assumed that the SWES will use OWL (Web Ontology Languages) ontologies, found in the Web, to generate rules, to supplement the SWES knowledge with these rules, and to reason, based on the rules from the SWES knowledge base and user interaction. The tasks of OWL ontology merging [4], rule generation from OWL ontologies [5], [6] as well as the task of the Jena framework adaptation for fuzzy reasoning had already been investigated and realized. It is expected that potential of the SWES will surpass potential of existing keyword-based search engines, when the task of web page transformation to OWL ontology will be investigated and solved. This task is for the future research.

The main purpose of this paper is to develop Ontology Web Search Engine project. This project will describe a search engine for searching and indexing OWL ontologies in the Web. The project is realizable to be integrated in the SWES. Implemented and integrated in the SWES Ontology Web Search Engine would allow the user to utilize all available OWL ontologies in the Web. Thus, the SWES would use knowledge from the Web.

This paper is divided into sections as follows. The next section gives an overview of existing ontology search opportunities in the Web. Section III presents several ontology search strategies in the Web. The following section describes the project of the Ontology Web Search Engine, which will be implemented in the near future. The last section presents the conclusions of this work.

II. RELATED WORK

The term "Semantic Web" was coined by Tim Berners-Lee more than 10 years ago [7]. Since then, this term came into use, and rapidly filled with content. Throughout this time the Semantic Web technologies were actively developed, that was why it was not surprising that the task of ontology search in the Web had already been studied and even implemented. There are several implemented ontology web search engines and let us look at them one by one.

SWOOGLE was the first search engine for the Semantic Web, and the main contributor of SWOOGLE was Li Ding [8]. SWOOGLE is positioned as a crawler-based indexing and retrieval system for the Semantic Web documents including RDF (Resource Description Framework) and OWL [9]. SWOOGLE consists of four major components SWD (Semantic Web Document) discovery, metadata creation, data analysis and interface. The SWD discovery component is responsible to discover the potential SWDs throughout the Web and keep up-to-date information about SWDs. The metadata creation component caches a snapshot of a SWD and generates objective metadata about SWDs in both syntax level and semantic level. The data analysis component uses the cached SWDs and the created metadata to derive analytical reports. The interface component focuses on providing data service to the Semantic Web community. The architecture of SWOOGLE is data centric and extensible: different components work on different tasks independently [9]. SWOOGLE is realized as a web page, and it is available at http://swoogle.umbc.edu. SWOOGLE is used by means of querying with keywords, after which the SWDs matching those keywords are returned in ranked order.

WATSON is one more search engine for the Semantic Web [10]. This search engine is available at http://watson.kmi.open.ac.uk/WatsonWUI/. WATSON is called the gateway for the Semantic Web, which has been guided by the requirements of Semantic Web applications and by lessons learnt from previous systems [11]. The goal of this gateway is to provide an efficient access point to the online ontologies and semantic data. WATSON collects the available semantic content on the Web, analyzes it to extract useful metadata and indexes and also implements efficient query facilities to access the data [11]. In order to support these three tasks, WATSON has been designed around three core activities, each corresponding to a "layer" of its architecture. These layers are the following [12]:
- The ontology crawling and discovery layer; it collects the online available semantic content by exploring ontology-based links.
- The validation and analysis layer; it is core to the architecture and ensures that data about the quality of the collected semantic information is computed, stored and indexed.
- The query and navigation layer; it grants access to the indexed data through a variety of mechanisms that allow exploring its various semantic features.

WATSON is similar to usual web or desktop search systems in the sense that it is based on keyword search [13]. In this regard, WATSON and SWOOGLE are alike.

A number of other Semantic Web search engines are known. Among them are such search engines as Falcons, Sindice, Semantic Web Search, SWSE (Semantic Web Search Engine) and others [10]. But they will not be discussed here for several reasons. Firstly, some of them are not available and cannot be tested (Falcons, Semantic Web Search, SWSE). Secondly, other Semantic Web search engines as Sindice do not work as effectively as it is necessary.

III. ONTOLOGY WEB SEARCH STRATEGIES

OWL ontologies are the main resource for the functioning of the SWES [5], [6]. In principle there is no difference how ontologies are supplied to the SWES. The main thing is that they would be, and it is desirable that they would be different. Different ontologies mean here that they are from different domains. OWL ontology web search strategies can be divided into the following categories:

- manual,
- automatic non-independent,
- automatic independent.

Manual OWL ontology web search strategy implies searching OWL ontologies using existing web search engines such as GOOGLE, YANDEX, SWOOGLE, WATSON and so on. There is no difference, what kind of a web search engine (a web search engine or a Semantic Web search engine) is utilized to implement this strategy. The uses of the Semantic Web search engines are preferable, because such systems are specially aimed at finding ontologies unlike the uses of the Web search engines. The use of existing web search engines gives an opportunity to get OWL ontology URLs (Uniform Resource Locator), to browse these ontologies and to download them, if it is necessary.

Automatic non-independent OWL ontology web search strategy implies searching OWL ontologies using built-in services of existing web search engines. SWOOGLE engine provides search services using REST (Representational State Transfer) interface [14]. It is possible to compose the query in an HTTP (Hypertext Transfer Protocol) GET query and retrieve the result as a dynamic web page encoded in RDF/XML. In turn, WATSON engine provides two types of open web services/APIs [15]:

- The Java/SOAP API, a complete Java API based on a number of SOAP/WSDL (Simple Object Access Protocol/ Web Services Description Language) services, providing complex search, querying and exploration mechanisms.
- The REST API, which provides a subset of the functionalities of WATSON through simple HTTP-based access, and giving back results in XML or JSON (JavaScript Object Notation), so that it can be, used easily in any language in particular Javascript.

WATSON provides three Web SOAP-based services to allow developers to programmatically access the semantic content (semantic data, ontologies) [16]. These web services have the associated API, which allows developing lightweight Semantic Web application. The WATSON REST API corresponds to a set of services simply accessible through HTTP calls [17]. For instance, it is possible to type certain URL in the Web browser to obtain the URIs (Uniform Resource Identifier) of all the semantic documents containing specific words.

Automatic independent OWL ontology web search strategy implies developing your own ontology web search engine. This strategy is the most difficult, but it has its own advantages. Of course, the SWES can be tested by means of several ontologies, which are manually found in the Web, but such a Semantic Web Expert System would not be complete. Having its own ontology web search engine, transforms the SWES from a research project into the system that is useful for the end user, and this is the first advantage of the strategy. External ontology web search services are always less reliable than exactly the same in terms of functionality, but built-in services. This is so, because any external services do not depend on our will, but the will of third-party developers is unpredictable, especially in the long term. External to the SWES ontology web search services, provided by SWOOGLE and WATSON, are not an exception. Thus, the SWES reliability increase, achieved by means of the SWES part reliability increase that is the SWES reliability increase, achieved by means of its own OWL ontology web search engine development is the second advantage of the strategy. The third advantage is that own developments are more modifiable, which is essential, if the SWES will continue to evolve. So, automatic independent OWL ontology web search strategy is preferable and therefore it will be implemented in Ontology Web Search Engine project, which will be described in the next section in detail.

IV. ONTOLOGY WEB SEARCH ENGINE PROJECT

Search engine technology keeps up with the growth of the Web. In 2000 there were more than 3200 search engines in the Web [18]. The number of documents, indexed by web search engines, is increasing from year to year. If WWWW (World Wide Web Worm) had an index of 110,000 web pages in 1994, then in 1997 search engines claimed to index from 2 million to 100 million web documents [19]. Nowadays GOOGLE, which is leading web search engine, has an index of over 30 trillion web pages [20]. Despite the differences of web search engines, the typical structure of a web search engine comprises four essential modules [18]:

- crawler,
- an indexer,
- a query engine,
- a page repository.

Crawlers are computer programs that browse the Web [18]. They use a starting set of URLs and retrieve (that is copy and store) the content on the Web pages specified by the URLs. The crawlers extract URLs appearing in the retrieved web pages and visit some or all of these URLs, thus repeating the retrieval process.

The indexer takes all the words from each document in the page repository and records the URL, where each word occurred [18]. The result is a very large database, which provides the URLs that point to pages, where a given word occurs. The database may also contain other structural information such as links between documents, incoming URLs to these documents, formatting aspects of the documents, and location of terms with respect to other terms .

The web query engine receives the search requests from users [18]. It takes the query submitted by the user, splits the query into terms, and also searches these terms in the database, which is built by the indexer. The web query engine then retrieves the documents that match the terms within the query and returns these documents to the user .

The page repository stores the Web content that was retrieved during the crawling process [18].

The web search engine typically ranks documents before presenting them to the user. The ranking process is calculating a similarity score between the query and each document. The higher the similarity score, the higher the ranking for a particular document [18].

Semantic Web search engines are very similar to traditional web search engines. For example, such a Semantic Web search engine as WATSON performs the same activities as traditional web search engines [13]:

- collecting the Semantic Web content,
- extracting useful metadata and indexing it,
- implementing efficient query facilities to access the metadata.

On this basis, it is possible to state that the main difference between traditional web search engines and Semantic Web search engines is in the content they work with. If traditional web search engines work primarily with documents in HTML, then Semantic web search engines work with semantic metadata, mainly contained in RDF and OWL documents. Otherwise, these web search engines are completely identical. Hence, it is necessary to develop its own crawler, indexer, query engine and page repository (ontology repository) for the OWL Ontology Web Search Engine.

The crawler for the Ontology Web Search Engine has to browse the Web to find OWL ontology URL's. There are two main strategies to accomplish the task of OWL ontology URL finding in the Web. The first strategy is to browse OWL ontologies from the Web to find the URL's of other OWL ontologies. This strategy is premature, because analysis shows that ontologies are not so common and not so qualitatively designed to use them for this task. However it should be noted that this strategy may be auxiliary or may be reserved for future use. The second strategy is to browse HTML pages from the Web to find the URL's of OWL ontologies and to find the URLs of other HTML pages to continue browsing the Web. This strategy is preferred for use, because analysis shows that the URL's of ontologies are basically presented on the web pages. The fact that web pages contain the URL's of other web pages has been known for a long time. There are two ways to implement the crawler for the Ontology Web Search Engine. The first way is to develop such a crawler using one of programming languages. The second one is to adapt one of existing crawlers to our needs. Existing crawlers may be divided into open sources crawlers and others. WebEater, Heritrix, JSpider, Java Web Crawler, Web-Harvest, Crawler4j, Nutch and so on are examples of open sources crawlers. It is necessary to identify crawler requirements in order to choose the way of crawler future implementation in the Ontology Web Search Engine. So, it would be preferable that the crawler would be written in Java, because the SWES is developing in Java. This crawler should have advanced functionality to be able to set the initial URL, from which the web crawling will start, to set the number of crawling web pages, to set the depth of crawling, to set the politeness of crawling that is how fast the crawler will apply to the new URL. Of course, if existing crawler is chosen, then it should be well documented.

It is necessary to utilize the tool for work with OWL ontologies in order to realize the indexer for the Ontology Web Search Engine. Jena [21] can serve as such a tool. This choice is due to several reasons. Firstly, Jena is an open source Semantic Web framework. Secondly, Jena is utilized by means of Java programming language, and this programming language is chosen for the SWES. Thirdly, Jena is used in other SWES subsystems, and it works well. Finally and this is the main reason, Jena gives an opportunity to access to all elements (classes, properties, relations and so on) of the OWL ontology. This opportunity allows storing the information of the ontology elements with the purpose to access to them later. Except the tool for work with OWL ontologies, it is necessary to realize something like database, which would have all necessary information about indexed ontology, in order to implement the indexer. There are two ways for this purpose. The first way is to develop such a database manually. The second way is to exploit one of existing indexers. Lucene, Nutch, Solr, Sphinx are just some of them. The task of developing or choosing the indexer for the Ontology Web Search Engine is the task for the future, however at the same time it is necessary to think about a query engine, performing this task.

The query engine is a subsystem of the Ontology Web Search Engine for searching the OWL ontology URLs, based on the user query. This means that the user types several keywords and the query engine returns the OWL ontology URLs, where the typed keywords appear. The query engine is closely related to the indexer, and actually the query engine looks for the ontology URLs in the database that is supplemented by the indexer.

The page repository of the Ontology Web Search Engine is storage, where the indexed OWL ontologies are located. Physically this storage will be a directory in the computer hard disc, where OWL ontologies will be collected. Such storage will give an opportunity to access to needed OWL ontology as quickly, as necessary.

Speaking about the whole structure of the Ontology Web Search Engine, it is necessary to note that this structure represents the set of the Ontology Web Search Engine functional parts, which are programmatically independent from each other. This means that there are several separate computer programs, which execute different tasks in the area of OWL ontology search in the Web. Thus, the structure of the Ontology Web Search Engine may look like this (Fig.1.):

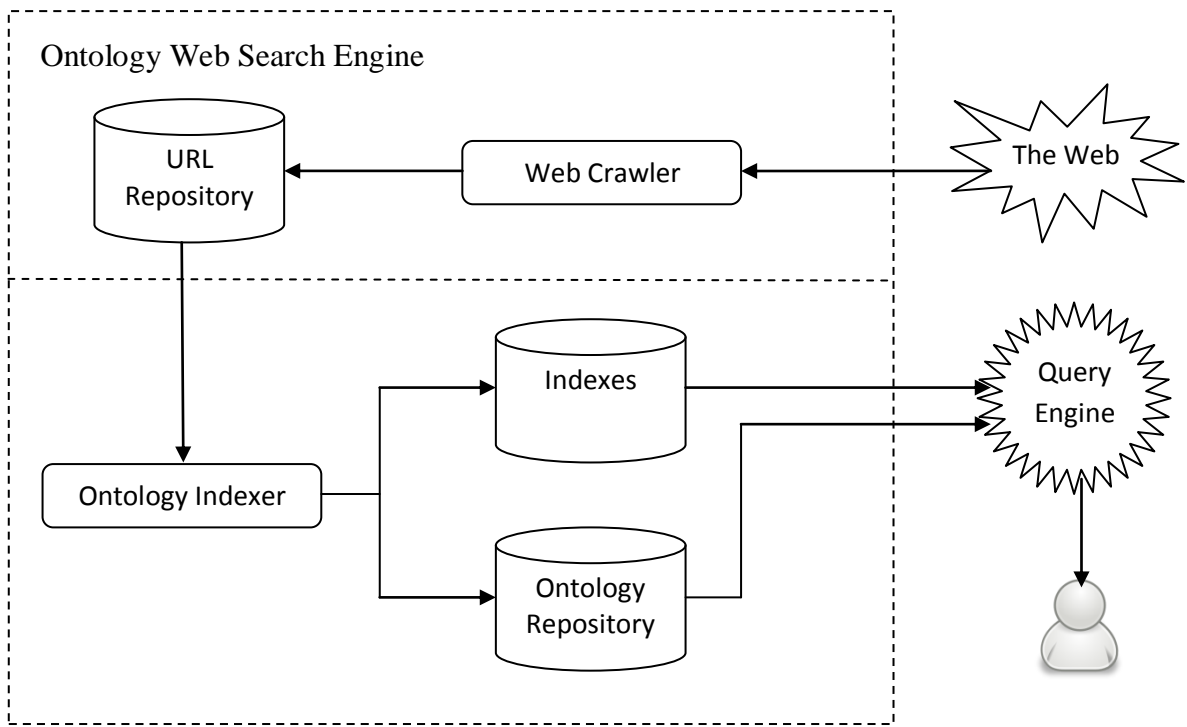

Fig. 1. The structure of the Ontology Web Search Engine.

The system of Ontology Web Search Engine consists of three computer programs ("Web Crawler", Ontology Indexer", "Query Engine") and three storages of information ("URL Repository", "Indexes", "Ontology Repository"). "Web Crawler" and "Ontology Indexer" are independent computer programs, but "Query Engine" is a software module that will be the part of the Semantic Web Expert System. "URL Repository" will contain the URL's of the OWL ontologies. "Ontology Repository" will contain OWL ontologies, and "Indexes" will contain the information about found OWL ontologies namely URL, size, classes, properties and relations (Fig. 1).

"Query Engine" (as the part of the SWES) is started, when "Indexes" and "Ontology Repository" storages are filled with the data. This occurs, when "Web Crawler" and "Ontology indexer" have already worked out. So, "Web Crawler" should be run first. It browses the Web in order to find the URL's of OWL ontologies. The URL's are stored in the "URL Repository", whenever they are found. "Web Crawler" will run until it finds a number of ontologies or it browses a number of web pages, as defined by the user. After the "Web Crawler" finishes running and the "URL Repository" is filled with the URLs, it is possible to run the "Ontology Indexer". The "Ontology Indexer" takes the URL's from the "URL Repository" in order, downloads the ontology at this URL, extracts the ontology components (properties, classes, relations), stores these information in the "Indexes" repository and also stores the ontology entirely in the "Ontology Repository". The "Ontology Indexer" runs until it processes all the URL's from the "URL Repository". The "Query Engine" software module works as follows. It receives the keywords, typed by the user, and looks for the matches in the "Indexes" repository. After the matches are found, the OWL ontology and its URL are identified. This is necessary for the further work of the SWES. In turn, the SWES will merge found ontologies, forming single one-domain ontology. Rules will be generated from this single ontology, and they will be used for reasoning, based on communication with the user.

V. CONCLUSION

The paper has described the project of Ontology Web Search Engine. The project aims to develop a Web Search Engine for searching and indexing OWL ontologies in the Web. This is necessary for the Semantic Web Expert System, which is expected to use OWL ontologies for its work [3]. In the course of the project description, the achievements of search engine technology have been overviewed, the typical structure of a web search engine has been described, ontology web search strategies has been identified, related works in the area of ontology search have been overviewed and also the structure of the Ontology Web Search Engine has been presented. This structure has been described in detail in order to have an opportunity to realize the project as soon, as it becomes possible.

It seems that OWL ontologies cannot displace regular web pages completely. At least it can be argued that this does not happen in the near future. Indeed, this is really difficult task, taking into account the distributed nature of the Web. Such a situation is in contradiction with the stated purpose of the SWES that is to be a universal expert system, which can use all the knowledge of the Web. Therefore it is necessary to be able to use the knowledge, generated from regular We pages, to remove this contradiction. One way to do this is to learn how to transform a regular web page, which, for example, is created in HTML, to OWL ontology. In turn, the process of knowledge (rule) generation from OWL ontology is a technical task, which has already been developed in [5], [6], [22]. So, the transformation of a regular web page to OWL ontology is one of the tasks to be developed.

REFERENCES

[1] J. Davis, R. Studer, P. Warren, "Semantic Web Technologies Trends and Research in On-tology-based Systems," John Wiley & Sons Ltd, Chichester, 2006.
[2] G. Antoniou, F. van Harmelen, "A Semantic Web Primer," 2nd ed., The MIT Press, 2008.
[3] O. Verhodubs and J. Grundspenkis, "Towards the Semantic Web Expert System", RTU Press, Riga, 2011.
[4] O. Verhodubs and J. Grundspenkis, „Ontology merging in the Context of the Semantic Web Expert System", Springer, Saint-Petersburg, 2013.
[5] O. Verhodubs and J. Grundspenkis, "Evolution of ontology potential for rule generation", Proceedings of the 2nd International Conference on Web Intelligence, Mining and Semantics, Craiova, 2012.

[6] O. Verhodubs, "Ontology as a Source for Rule Generation", 2014.
[7] T. Berners-Lee, J. Hendler and O. Lassila, "The Semantic Web", 2001.
[8] http://ebiquity.umbc.edu/person/html/Li/Ding/
[9] L. Ding, T. Finn and others, "Swoogle: A Search and Metadata Engine for the Semantic Web", 2004.
[10] http://www.w3.org/wiki/TaskForces/CommunityProjects/LinkingOpenData/SemanticWebSearchEngines
[11] http://watson.kmi.open.ac.uk/Overview.html
[12] M. d'Aquin, C. Baldassarre and others, „WATSON: Supporting Next Generation Semantic Web Applications", 2007.
[13] M. d'Aquin, E. Motta, „Watson, more than a Semantic Web search engine", 2011.
[14] http://swoogle.umbc.edu/index.php?option=com_swoogle_manual&manual=search_overview#Search%20ontology
[15] http://watson.kmi.open.ac.uk/WatsonWUI/
[16] http://watson.kmi.open.ac.uk/WS_and_API-v2.html
[17] http://watson.kmi.open.ac.uk/REST_API.html
[18] A. Spink and B. Jansen, „WEB SEARCH: PUBLIC SEARCHING OF THE WEB", 2004.
[19] S. Brin and L. Page, „The Anatomy of a Large-Scale Hypertextual Web Search Engine", 1998.
[20] http://searchengineland.com/google-search-press-129925
[21] https://jena.apache.org/
[22] O. Verhodubs, „Inductive Learning for Rule Generation from Ontology", 2015.
[23] O. Verhodubs, „Adaptation of the Jena framework for fuzzy reasoning", 2014.